\documentclass[showpacs,prb,aps,twocolumn]{revtex4}
\usepackage{graphicx}
\usepackage{color}
\usepackage{dcolumn}
\usepackage{bm}
\usepackage{amsmath}
\def\n{\mathbf{n}}
\def\m{\mathbf{m}}
\def\k{\mathbf{k}}
\def\gapx{\lower 2pt \hbox{$\buildrel>\over{\scriptstyle{\sim}}$\ }}
\def\lapx{\lower 2pt \hbox{$\buildrel<\over{\scriptstyle{\sim}}$\ }}
\begin{document}
\title{Variational study of a mobile hole in a two-dimensional quantum antiferromagnet using entangled-plaquette states}

\author{Fabio Mezzacapo}

\affiliation{Max-Planck-Institut f\"ur Quantenoptik, Hans-Kopfermann-Str.1, D-85748, Garching, Germany}
\date{\today}

\begin{abstract}
We study the properties of a  mobile hole in the $t-J$  model on the square lattice by means of variational Monte Carlo simulations based on the entangled-plaquette ansatz. Our energy estimates for small lattices  reproduce  available exact results. We obtain
values for the hole energy dispersion curve on large lattices in quantitative agreement with earlier findings based on the most reliable numerical techniques. Accurate estimates of the hole spectral weight are provided.
 \end{abstract}


\maketitle

\section{Introduction} 
Since the fundamental suggestion made by Anderson in 1987, the two-dimensional (2D) $t-J$ model ($tJ$M) has been regarded as an effective hamiltonian description of the basic properties of the superconducting  copper oxides.\cite{And87, Z88} Essential physical features of the insulating copper-oxide planes of  these compounds  are successfully reproduced by the spin-$\frac{1}{2}$ antiferromagnetic Heisenberg model (AHM),\cite{Man91} to which the $tJ$M reduces when holes are absent.  On  the other hand, such insulating  planes can turn superconducting if doped with mobile holes,  whose presence is accounted for in the $tJ$M by adding to the AHM a nearest-neighbor hole-hopping term.  Hence, the $t-J$ Hamiltonian is
\begin{equation}
H=-t\sum_{\langle i,j \rangle,s}(\overline{c}^+_{i,s}\overline{c}_{j,s}+h.c.)+J\sum_{\langle i,j \rangle}(\mathbf{S}_i \cdot \mathbf{S}_j - \frac{1}{4}\hat{n}_i\hat {n}_j)
\label{eq:ham}
\end{equation}
where the summations run over nearest-neighbor sites on the square lattice, $t$ is the hole hopping amplitude, and $J$ the antiferromagnetic coupling between site $i$ and $j$; $\overline{c}^+_{i,s}$ creates an electron, at site $i$, of spin projection (e.g.,  along the $z$ axis) $s=\pm\frac{1}{2}$ avoiding double occupancy. Therefore, in  terms of the usual fermionic creation operator  $\overline{c}^+_{i,s}=\hat{c}^+_{i,s}(1- \hat{n}_{i,-s})$; $\mathbf{S}_i$ and  $\hat{n}_i=\hat{n}_{i,s}+\hat{n}_{i,-s}$, with $\hat{n}_{i,\alpha}=\hat{c}^+_{i,\alpha}\hat{c}_{i, \alpha}$, are the spin-$\frac{1}{2}$  and  the number operator respectively. 

Despite the simple form of the above Hamiltonian, its study constitutes, even in presence of a single hole, a very difficult problem. Exact diagonalization (ED), possible only for small system sizes (i.e., up to $N\sim26$ sites),\cite{Dag91, Bar92, Poi93} approximated analytical treatments,\cite{Kan89, Mar91, Man91a, Hor91} as well as numerical approaches  of various type,\cite{Mas91, Mas92, Mas94, Mur00} have been largely employed to gain insight into the physics of the one-hole $tJ$M.  
Specifically, quantum Monte Carlo (QMC) techniques based on imaginary time projection furnish, even for large systems,  accurate (exact in principle) results at half filling (i.e., no-hole case) or for $t$=0 (i.e., static-hole case).\cite{Mas94,Mur00} However, when $t \neq 0$ these methods are affected by the sign problem, for which only essentially uncontrolled workarounds (e.g., fixed-node approximation) exist. Furthermore, being  a 2D one, the model of our interest cannot be tackled, due to the unfavorable scaling of the required computational resources, by using the Density matrix renormalization group (DMRG),\cite{Whi92} the accepted method of choice  for 1D lattice models. Variational Monte Carlo (VMC) schemes  do not suffer from the sign problem; on the other hand they are not strictly exact, their accuracy  depending  on the particular choice of the trial wave function (WF). For the $tJ$M with a single hole, specific variational WF's \cite{Mas91, Mas92} have been designed affording estimates  of the hole dispersion relation in reasonable agreement with those, generally more accurate, obtained by QMC. 

Recently, the suggestion was made of combining DMRG and Monte Carlo in order to design novel variational algorithms for tensor-network based ansatze applicable to systems of spatial dimension higher than one.\cite{San07, Sch0810} The WF introduced by us,\cite{eps} founded on entangled-plaquette states (EPS), has been successfully employed to  study  ground-state (GS) properties of lattice spin models with open or periodic boundary conditions (PBC). Such ansatz has been proven to yield  estimates of physical observables of accuracy at least comparable to that obtainable with the best alternative techniques or trial WF's.\cite{eps, cps, eps1,cps1} The EPS WF can be efficiently optimized for  plaquettes of a given size, as well as  systematically improved by adding more plaquettes and/or increasing their size.\cite{eps,cps}

In this paper we apply the EPS ansatz to investigate the dynamics of a hole in a quantum antiferromagnet described in Eq.  (\ref{eq:ham}). Specifically, we compute the hole energy dispersion curve $E(\k)$, as well as the spectral weight of the quasi-hole state, for lattice sizes up to $N = 256$ sites in the parameter-range $0.4 \le \frac{J}{t} \le 2.0$ (the $tJ$M  is believed to be relevant to the cuprate superconductors for a value of $\frac{J}{t} \sim 0.4$). From the methodological standpoint, this study is considerably more challenging than that of the various (un)frustrated spin models to which the EPS ansatz has been recently applied with success.\cite{eps, eps1}

The main goal of this work is to demonstrate the accuracy of the EPS WF in describing the property of a mobile hole in the $tJ$M. This is achieved by a systematic comparison of our results for the energy dispersion curve and the hole spectral weight with estimates computed by  different approaches.   Furthermore, besides the mentioned methodological aspects we  offer extremely accurate results for the hole spectral weight in the so called ``physical region'' (i.e.,$ \frac{J}{t}\sim 0.4$) where the accuracy of  QMC data, the most reliable ones obtained so far, becomes less satisfactory.\cite{Mur00}
Assessing the reliability of the EPS ansatz in a case (i.e., the $tJ$M with one mobile hole)  for which a large number of results obtained with alternative trial states, or other  methodologies are available in literature, is a fundamental step also in view of future applications of our WF  to the long-standing problem of understanding the phase diagram of the $tJ$M at finite hole concentration. The accuracy of the estimates presented here is clearly not a proof of the reliability of our variational choice  when more holes are present, it constitutes, however, the  necessary condition to consider the EPS WF, with no need of essential adjustments, a promising  option for the investigation of  the finite hole concentration scenario. A direct study of the latter problem, given the paucity of reliable results obtained in the past, would render the assessment of the trustability of our ansatz problematic leaving, on the other hand, room for dubitative, definitely well-posed,  questions such as: {\it Why results at finite hole concentration should be taken seriously if the  reliability of the EPS WF has not even been proved in the simpler  single hole case ?}

On using a minimal EPS ansatz based on $N$ $2\times 2$ entangled plaquettes, we obtain estimates of the hole dispersion curve in qualitative and semi-quantitative agreement with those computed by QMC. We achieve quantitative agreement, also for other observables, with an EPS WF  based on plaquettes comprising  $9$ lattice sites. 

The accuracy of our findings for  one hole  confirms the flexibility of the EPS ansatz, which can be  applied with no substantial modifications to a large class of  lattice models. More importantly, the EPS WF used in this work is easily generalizable  to the case of a finite hole concentration, where the sign problem renders QMC inapplicable, and the generalization of different trial states employed for  one or two holes  appears quite complicated.\cite{Mas91, Mas92}  

\section{Variational wave function}
Let us consider a  square lattice of $N$ sites  comprising one  electron per site,  a generic WF can be written as
\begin{equation}
|\Phi\rangle=\sum_{\mathbf{n}}W(\mathbf{n})|\mathbf{n} \rangle
\end{equation}

where  $|\mathbf{n} \rangle = |n_1,n_2, \ldots , n_N\rangle$,  $n_i $ is the eigenvalue of $
\sigma^z_i $ and $W(\mathbf{n})$  is the weight of a configuration of the system. 
The EPS ansatz is constructed as follows:

i) Cover the systems with $M$ plaquettes in such a way that $l$ sites labelled by $n_{1,P},n_{2,P},\ldots,n_{l,P}=\n_P$  belong to the $P_{th}$ plaquette.

 ii) Express the weight of a global spin configuration as a product of variational coefficients $C_1^{\n_1},C_2^{\n_2},\ldots,C_M^{\n_M}$   in biunivocal correspondence to the particular configuration  (e.g., spin state along the $z$ axis of the plaquette sites)  of the plaquettes. Hence, taking into account explicitly the bipartite nature of the spin-$\frac{1}{2}$ square Heisenberg antiferromagnet at half filling:   \begin{equation}
\langle \n | \Phi \rangle=W(\n)=(-1)^{F(\n)}\mathcal{P}(\n)
\label{coeffh}
\end{equation}
being $F(\n)$, according to the Marshall-sign rule, the number of  ``down'' spins in one of the two sublattices\cite{Mar55} and
\begin{equation}
\mathcal{P}(\n)=\prod_{P=1}^M C_P^{\n_{P}}
\end{equation}
the crucial  part of the ansatz.

Analogously, in presence of a single hole, our EPS WF is defined via
\begin{equation}
\langle \m | \psi_{\mathbf{k}} \rangle=W(\m)=
e^{-i\mathbf{k}\mathbf{R}_h}(-1)^{F(\m)}\mathcal{P}(\m);
\label{coeffk}
\end{equation}
here $\m$ refers to the generic configuration containing one hole and $N-1$ electrons,
 $\mathbf{k}$ and $\mathbf{R}_h$ are the hole momentum and the spatial position of the empty site respectively.
 It is worth mentioning that the trial states adopted in this work naturally incorporate, for plaquettes including more than one site,   electron-electron as well as electron-hole correlations.
We estimate GS energies of the Hamiltonian in Eq. (\ref{eq:ham}) via variational optimization of the states (\ref{coeffh}) at half filling and (\ref{coeffk}) in presence of a single hole. Independent optimizations are carried on for each $\mathbf{k}$ value by Monte Carlo sampling  both the energy and its derivatives with respect to the varational parameters (i.e., the $C^{\n_{P}}_P$'s). Further details concerning the variational family of EPS and the numerical technique adopted to minimize the energy are given in Refs. \onlinecite{eps} and \onlinecite{eps1}.
Although the EPS ansatz is exact when a single plaquette as large as the system is employed,  accurate results are already obtainable by using $N$ entangled (i.e., overlapping) plaquettes comprising $4$ sites, the accuracy of the estimates being sensibly improvable increasing the plaquette size.
Specifically, we compute the hole energy dispersion curve 
\begin{equation}
E(\mathbf{k})=E^{tJ\text{M}}(\mathbf{k})-E^{\text{AHM}},
\label{eq:disp}
\end{equation} 
defined by the difference between the $\k$-dependent GS energy of the one-hole $tJ$M and that of the AHM, and the hole spectral weight
\begin{equation}
Z(\mathbf{k})=|\langle\psi_{\mathbf{k}}|\overline{c}_{\mathbf{k}}|\Phi\rangle|^2,
\label{eq:zeta}
\end{equation}
where $|\psi_{\mathbf{k}}\rangle$ ($|\Phi\rangle$) is the normalized GS of the one (zero) -hole system.

We compare the EPS results with exact ones for a $4\times4$ lattice and  with the most accurate available in literature, for larger systems. In this case the main features of the hole dispersion relation, namely the characteristic shape with a  global minimum at $\mathbf{k}=(\pm\frac{\pi}{2},\pm\frac{\pi}{2})$ and a  nearly flat region around $\mathbf{k}=(\pm \pi,0)$, are very well reproduced with an ansatz based on $2\times2$ plaquettes, while the increasing of the plaquette size to $3\times3$ gives rise to an approximately  rigid shift, towards lower energies, of the band and an almost perfect  agreement (not reachable, to our knowledge, with other variational WF's) with  previous QMC studies for all the $\frac{J}{t}$ values considered in this work.  Moreover, with the $3\times3$ EPS,  accurate estimates of the hole spectral weight are obtained, on the same footing, via the Monte Carlo method.

\section{Results}
The GS energies of the one-hole $tJ$M on a $4\times4$ lattice can be obtained, given the small system size, by using a single plaquette comprising all the lattice sites. In this case the EPS ansatz is exact, the error bar of our estimates being only due to the limited number of configurations sampled, and reproduces, as expected, ED results.\cite{Bar92} 

Clearly, for larger systems, the use of a single plaquette is not feasible, therefore, it is important to asses the accuracy of our WF for smaller plaquette size. The error relative to the exact result: $E^{tJ\text{M}}(\pm\frac{\pi}{2},\pm\frac{\pi}{2})=-21.161$ of our estimate  of the global energy minimum at $\frac{J}{t}=0.5$  is  $\sim 3$\% for the $2\times2$ EPS ansatz, decreasing down to less than $0.2$\% when $3\times3$ plaquettes are used. For this plaquette size, our results compare favorably to those obtained with different trial states, even when larger lattices are considered. For example, the EPS   energy when $\mathbf{k}=(\pm\frac{\pi}{2},\pm\frac{\pi}{2})$ on a $8\times8$ lattice at $\frac{J}{t}=0.4$ is $-78.463(7)J$,  while the WF including both spin-spin and momentum dependent spin-hole correlations adopted  in Ref. \onlinecite{Mas91}  yields a value of $-76.753(14)J$ (i.e.,  $\sim 2$\% higher).

 \begin{figure}[b] 
 \begin{center}
{\includegraphics[scale=0.7]{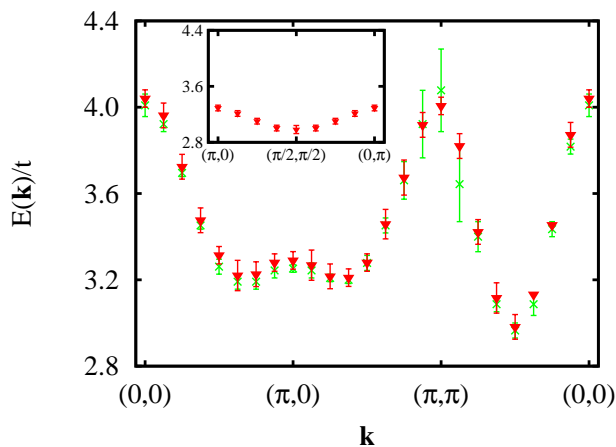}}
\end{center}
\caption{(color online). Hole energy dispersion curve for a lattice of $256$ sites with PBC at $\frac{J}{t}=2.0$. EPS estimates (triangles) are computed using square plaquettes of 9 sites. QMC results (crosses) are also shown for comparison.\cite{Mur00}}
\label{fig:1}
\end{figure}
\begin{figure}[t]
 \begin{center}
{\includegraphics[scale=0.7]{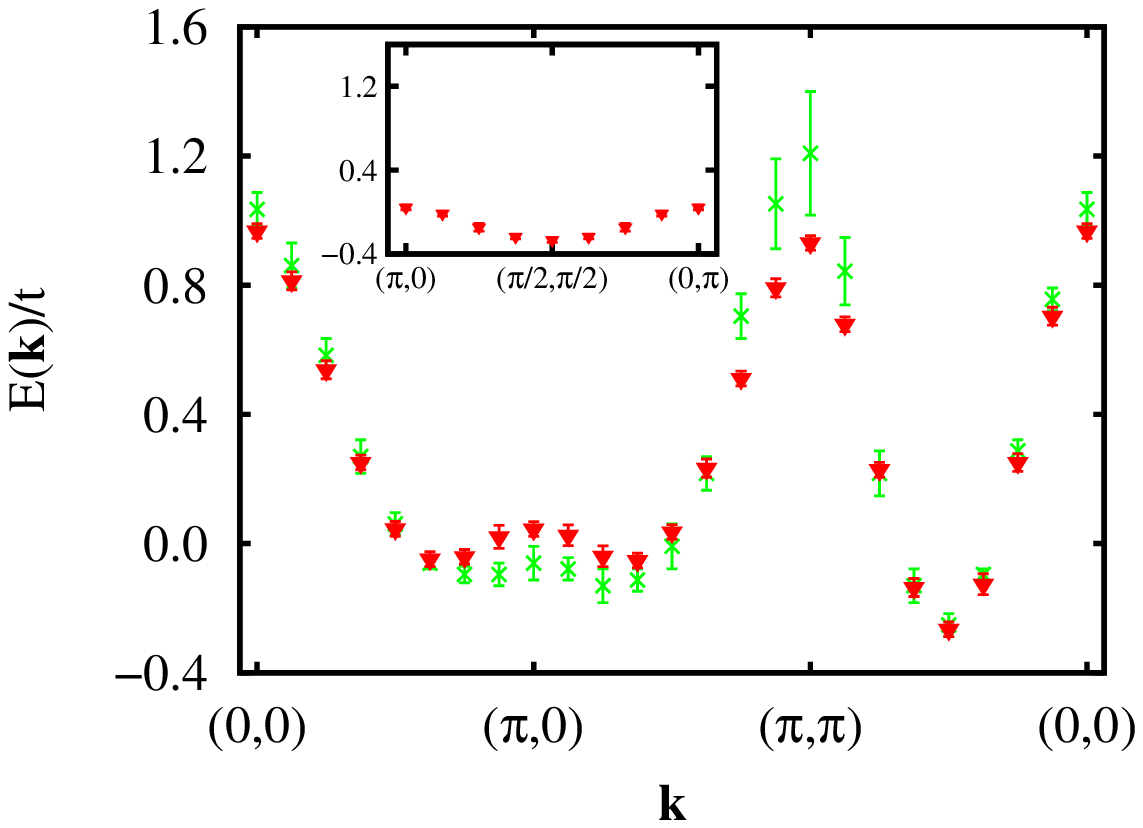}}
\end{center}
\caption{(color online). Hole energy dispersion curve for a lattice of $256$ sites with PBC at $\frac{J}{t}=0.8$. EPS estimates (triangles) are computed using square plaquettes of 9 sites. QMC results (crosses) are also shown for comparison.\cite{Mur00}}
\label{fig:2}
\end{figure}
\begin{figure}[b]
 \begin{center}
{\includegraphics[scale=0.7]{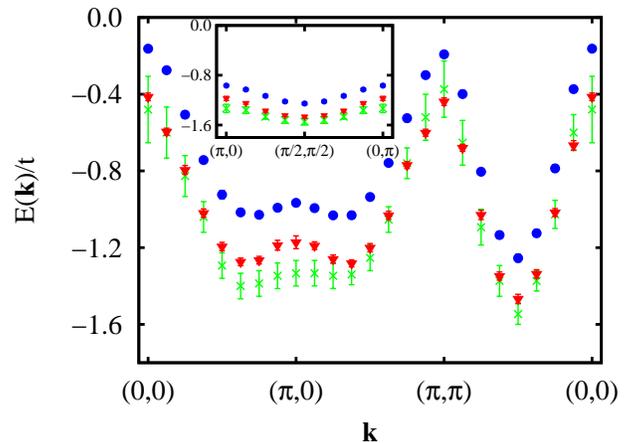}}
\end{center}
\caption{(color online). Hole energy dispersion curve for a lattice of $256$ sites with PBC at $\frac{J}{t}=0.4$. EPS estimates  are computed using square plaquettes of 4 (circles) or 9 (triangles) sites. GFMC results (crosses)\cite{Mas94} are also shown for comparison.} 
\label{fig:3}
\end{figure}

As opposed to the case in which more holes are present and the sign problem is particularly severe, QMC techniques based on imaginary time projection, for the one-hole $tJ$M, provide extremely accurate results, to our knowledge the most accurate  in literature, when, due to the large system size, the ED of the Hamiltonian is not possible. Hence, a comparison of the EPS WF against these approaches is in order.

The hole energy dispersion curve, defined in Eq. (\ref{eq:disp}), is shown in Fig. \ref{fig:1} for the $16\times16$ lattice at $\frac{J}{t}=2.0$. Our estimates (triangles), computed with the $3\times3$ EPS ansatz, are indistinguishable (taking into account the error bars) from available QMC ones\cite{Mur00} (crosses). Around $\k=(\pi,0)$ the curve is nearly flat, in agreement with other studies.\cite{Mas91, Ham98} It has to be mentioned that the degeneracy observed on the $4\times4$ lattice between $E(\pi,0)$ and $E(\frac{\pi}{2}, \frac{\pi}{2})$ is due to geometrical reasons\cite{Bar92} and is no longer present when the lattice size increases.

The hole dispersion relation for lower values of $\frac{J}{t}$ is plotted in Figs. \ref{fig:2}  and \ref{fig:3}. The main qualitative features of the curve (e.g., the position of the minimum) appear independent of the value of $\frac{J}{t}$. To  illustrate further the accuracy of the EPS WF, estimates of  $E(\k)/t$  obtained with $2\times2$ plaquettes (circles in Fig. \ref{fig:3}) are also shown for $\frac{J}{t} = 0.4$. Despite the relatively small number of variational coefficients utilized  for this ansatz, the resulting dispersion relation is in close agreement with that found with $3\times3$ plaquettes as well as with Green function Monte Carlo (GFMC) results\cite{Mas94}  (triangles and crosses respectively in the same figure) and QMC ones.\cite{Mur00} Specifically, the hole band predicted by using such a minimal EPS ansatz differs from the more accurate prediction  achievable with $3\times3$ plaquettes only by a rigid energy shift towards lower energy values. This energy shift (albeit quantitatively different)  occurs for all the three values of $\frac{J}{t}$ considered here, pointing out how the EPS ansatz, even in its simplest and  economical implementation (i.e., when $2\times2$ plaquettes are used), can provide reliable estimates of observable such as the effective mass, the band width, and generally all those obtainable from band energy differences.
 \begin{table}[t]
\caption{\label{tab:2}Band width ($W$) and energy difference ($G$) 
between $E(\pi,0)$ and $E(\frac{\pi}{2},\frac{\pi}{2})$ 
, in units of $J$, for a system of $256$ sites with PBC. 
Error bars are in parenthesis.}
\begin{ruledtabular}
\begin{tabular}{ccc}
$\frac{J}{t}$ & W & G  \\
\hline
2.0 & 0.53(3) & 0.15(3)  \\
0.8 & 1.54(3) & 0.39(3)  \\
0.4 & 2.63(4) & 0.74(8)  \\
\end{tabular}
\end{ruledtabular}
\end{table}
\begin{table}[b]
\caption{\label{tab:3}Hole spectral weight at $\k=(\frac{\pi}{2},\frac{\pi}{2})$ for a $16\times16$ lattice with PBC. Estimates  obtained with VMC based on a different WF,\cite{Mas92} QMC,\cite{Mur00} and SCBA\cite{Hor91} are also shown for comparison. 
Error bars are in parenthesis.}
\begin{ruledtabular}
\begin{tabular}{ccccc}
$\frac{J}{t}$ & EPS & VMC & QMC & SCBA  \\
\hline
2.0 & 0.596(2) & 0.663(3) & 0.58(4) & -  \\
0.8 & 0.427(3) & - &  0.40(4) & 0.504 \\ 
0.4 & 0.340(2) & 0.375(2) & 0.32(2) & 0.34 \\
\end{tabular}
\end{ruledtabular}
\end{table}

Values of the band width $W$ and the energy gap $G$ respectively defined as $W=E(0,0)-E(\frac{\pi}{2},\frac{\pi}{2})$ and $G=E(\pi,0)-E(\frac{\pi}{2},\frac{\pi}{2})$
 are reported, as a function of $\frac{J}{t}$, for a $16\times16$  lattice in Tab. \ref{tab:2}. Our results of the band width are in excellent agreement with QMC\cite{Mur00} and GFMC\cite{Mas94} calculations. At $\frac{J}{t}=2.0$ our estimate  differs from the VMC one: $2.11(3)$ reported  in Ref. \onlinecite{Mas92}. Moreover we find that $G$, at least in the cases examined here, varies linearly with $t$ according to the relation  $\frac{G}{t} \sim 0.3$.

Next we discuss our findings for the hole  spectral weight $Z$ (see Eq. (\ref{eq:zeta})). On a $4\times4$ lattice we obtain, for $\frac{J}{t}=0.4$, $Z(\frac{\pi}{2},\frac{\pi}{2})=0.3996(5)$, in agreement with the ED value\cite{Dag91} of $0.4$. EPS values of $Z(\frac{\pi}{2},\frac{\pi}{2})$ as a function of $\frac{J}{t}$ on a $16\times16$ lattice are shown in Tab. \ref{tab:3}. Our estimates, computed with plaquettes of $9$ sites, are in agreement with  QMC  ones\cite{Mur00} for all the values of $\frac{J}{t}$; agreement is found with self consistent Born approximation (SCBA) calculations\cite{Hor91} at $\frac{J}{t}=0.4$, this approach providing, as well as the WF used in Ref. \onlinecite{Mas92}, higher values of the spectral weight when $\frac{J}{t}$ increases. 

\begin{figure}[t]
 \begin{center}
{\includegraphics[scale=0.7]{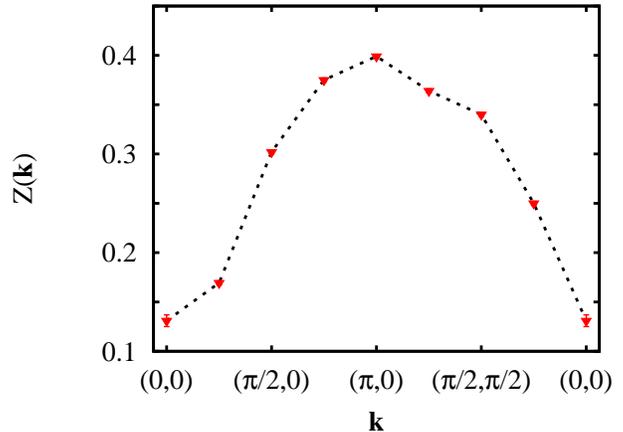}}
\end{center}
\caption{(color online). Hole spectral weight for a $16\times16$ lattice with PBC at $\frac{J}{t}=0.4$. The dashed line is a guide to the eye.}
\label{fig:4}
\end{figure}

The momentum dependence of the spectral weight for $\frac{J}{t}=0.4$ is shown in Fig. \ref{fig:4} along particular cuts of the Brillouin zone. $Z(\k)$  is approximately $13$\% at $\k=(0,0)$ and, going in the  $(1,0)$ direction,  reaches $\sim 0.4$ at $\k=(\pi,0)$; it then   decreases, slightly, down to $\sim 0.34$ at $\k=(\frac{\pi}{2},\frac{\pi}{2})$, and more pronouncedly, in the $(1,1)$  direction,   
beyond this $\k$ value. 
 
\section{Conclusions and outlook}
The two dimensional $t-J$ model in presence of a single mobile hole has been investigated by means of variational Monte Carlo simulations based on the EPS trial wave function. For systems comprising up to $N=256$ sites (i.e., the maximum size considered in this work), an EPS ansatz consisting of $N$ entangled plaquettes of $4$ sites provides estimates of the dispersion curve in qualitative and semi-quantitative agreement (up to a rigid energy shift) with the most accurate alternative numerical approaches. By increasing the plaquette size to $9$ sites, with a minimal additional computational effort, quantitative agreement is easily recovered and accurate results for the hole spectral weight obtained.

Given the accuracy of the results presented in this work,  as well as the simplicity of the variational algorithm used to minimize the energy, the EPS ansatz appears a very promising option for future investigations of the $t-J$ model in the case of finite hole concentration, a problem of great physical interest to which exact quantum Monte Carlo techniques based on imaginary time projection are not applicable due to the sign problem.
\section*{Acknowledgments}
The author acknowledges discussions with J. I. Cirac. 
This work has been supported by the DFG (FOR 635) and the EU project QUEVADIS.

\end{document}